\documentclass[12pt]{article}
\usepackage{amsmath}
\usepackage{graphicx}
\begin{document}

\title{Axisymmetric, Rotating and Stratified Stars}
                   
\author{Mayer Humi\\
Department of Mathematical Sciences\\
Worcester Polytechnic Institute\\
100 Institute Road\\
Worcester, MA  01609}

\maketitle
\thispagestyle{empty}

\begin{abstract}

The paper considers Euler-Poisson equations which govern the 
steady state of a self gravitating, rotating, axi-symmetric stars under the 
additional assumption that it is composed of incompressible stratified fluid. 
The original system of six nonlinear equations is reduced to 
two equations, one for the mass density and the other for gravitational field.
This reduction is carried out separately in cylindrical and spherical 
coordinates. As a "byproduct"  we derive also expressions for the pressure. 
The resulting equations are then solved approximately
and these analytic solutions are used then to determine the shape of the 
rotating star.
\end{abstract}

\newpage

\section{Introduction}

The steady states of self gravitating fluid in three dimensions 
have been studied by a long list of theoretical physicists and astrophysicists. 
(For an extensive list of references see [3,4,17,22,23]). In fact the research 
along these lines persists even today [9,10,15,16,20,21]. The motivation for this 
research is due to the interest in the formation, shape and stability of stars 
and other celestial bodies.  

Within the context of classical mechanics attempts to describe star interiors
are based on Euler-Poisson equations [3,4]. Well known solutions to these 
equations are the Lane-Emden functions which describe steady state 
non-rotating spherically symmetric stars with mass-density $\rho=\rho(r)$ 
and flow field $\bf u=0$.
The generalization of these equations to include axi-symmetric rotations was 
considered by Milne [18], Chandrasekhar [3,4] and many others [11,14,15,19,20]. 
One of  difficulties in the treatment of this problem is due to the fact that 
the boundary of the domain can not be prescribed apriori and one has to address 
a free boundary problem. An approximate treatment of this problem for 
polytropic stars in spherical coordinates was made in [21]. Another approach 
to this problem using the conservation of the total mass and angular 
momentum was initiated by Auchmuty and Beals [1] and was followed by many 
others [2,5,11,12,13,16].

In the present paper we address the modeling of axi-symmetric rotating stars
from a different perspective. Thus we add the assumption that the 
mass-density is stratified [6,7,17,22,23] to the Euler-Poisson equations 
with axi-symmetric rotations. Under these assumptions we show that the number 
of model equations for the steady state can be reduced from six to a system 
of two coupled equations. One for the mass-density and the second for the 
gravitational field. These equations contain, however, a parameter  
function $h(\rho)$ that encode the information about the momentum 
distribution within the star. This reduction in the number 
of model equations (for this class of stars) may be used to obtain new 
insights for the treatment of this problem and make it tractable both 
analytically and numerically. We provide in this paper approximate analytic
solutions to these equations and use these 
solutions to solve for the shape of the rotating star.

It might be argued that Euler-Poisson equations do not actually hold
in a star interior due to the various physical processes taking place
there (e.g turbulence, radiation, compressibility etc). Nevertheless they 
provide a natural extension to the results on the equilibrium states of 
three dimensional bodies under gravity.
  
The plan of the paper is as follows: In Sec 2 we present the basic 
model equations. In Sec. 3 we carry out in cylindrical coordinates
the reduction of these model equations from six to two.
We provide also expressions for the pressure in this coordinate system. 
In Sec 4 we discuss analytic solutions to these
equations and derive in one instance explicit expression for the shape 
of the rotating star. In Sec 5 we carry out the same reduction in spherical
coordinates. We end up in Sec 6 with summary and conclusions.

\setcounter{equation}{0}
\section{Derivation of the Model Equations}

In this paper we consider the state of an inviscid 
incompressible stratified self gravitating fluid. In addition we assume that 
the fluid it is subject to axial rotations. 
The hydrodynamic equations that govern this flow in an inertial frame of
reference are [2,3,6,7,19,22];
\begin{equation}
\label{2.1}
\nabla \cdot {\bf v} = 0
\end{equation}
\begin{equation}
\label{2.2}
{\bf v} \cdot \nabla\rho = 0
\end{equation}
\begin{equation}
\label{2.3}
\frac{1}{2}\rho\nabla({\bf v}\cdot{\bf v})+
\rho(\nabla\times{\bf v})\times {\bf v} = 
-\nabla p -\rho \nabla \Phi 
\end{equation}
\begin{equation}
\label{2.4}
\nabla^2 \Phi = 4 \pi G \rho
\end{equation}

where ${\bf v}=(u,v,w)$ is the fluid velocity, $\rho$ is its density 
$p$ is the pressure, $\Phi$ is the gravitational potential, G is 
the gravitational constant and the momentum equations (\ref{2.3}) are 
written in Lambs's form. Subscripts denote differentiation with respect
to the indicated variable.

We can nondimensionalize these equations by introducing the following scalings
\begin{eqnarray}
\label{2.5}
&&x= L\tilde{x},\,\,\ y=L\tilde{y},\,\,\, z=L\tilde{z},\,\,\, 
{\bf v}=U_0 \tilde{{\bf v}} ,\,\,\\ \notag
&&\rho = \rho_0 \tilde{\rho},\,\,\ p=\rho_0 U_0^2\tilde{p},\,\,\,
\Phi= U_0^2 \tilde{\Phi},\,\,\, \omega=\frac{U_0}{L}\tilde{\omega}.
\end{eqnarray}
where $L,U_0,\rho_0$ are some characteristic length,velocity and mass density
respectively that characterize the problem at hand.

Substituting these scalings in (\ref{2.1})-(\ref{2.4}) and dropping 
the tildes these equations remain unchanged (but the quantities that appear
in these equations become nondimensional) while $G$ is replaced by 
$\tilde{G}=\frac{G\rho_0 L^2}{U_0^2}$. (Once again we drop the tilde).

We now restrict our discussion to bodies which are axi-symmetric.
Without loss of generality we shall assume henceforth that this axis
of symmetry coincides with the z-axis. Under this assumption it is
expeditious to treat the flow either in cylindrical or spherical coordinate
system. In standard cylindrical coordinates $(r,\theta,z)$ 
we then have (due the symmetry) ${\bf v}= {\bf v}(r,z)$  i.e. the flow 
and the other functions that appear in (\ref{2.1})-(\ref{2.4}) are
independent of the angle $\theta$. 

\setcounter{equation}{0}
\section{Reduction in Cylindrical Coordinates}

Following the standard notation we introduce the frame
$$
{\bf e}_r =(\cos\theta,\sin\theta,0),\,\,\, {\bf e}_{\theta} =
(-sin\theta,\cos\theta,0),\,\,\, {\bf e}_z=(0,0,1).
$$
In this frame we have under present assumptions
\begin{equation}
\label{2.00}
{\bf v} = u(r,z){\bf e}_r + w(r,z){\bf e}_z+
v(r,z){\bf e}_{\theta}={\bf u}(r,z)+v(r,z){\bf e}_{\theta}
\end{equation}
The momentum equations for ${\bf u}$ can be written as
\begin{equation}
\label{2.02}
\rho{\bf u}\cdot \nabla {\bf u}=-\nabla p-
\rho\nabla\Phi+ \rho\frac{v^2}{r}{\bf e}_r.
\end{equation}
The equation for $v$ is
\begin{equation}
\label{2.03}
{\bf u}\cdot\nabla v+\frac{uv}{r}=0.
\end{equation}
We observe also that we can replace ${\bf v}$  by ${\bf u}$ in 
(\ref{2.1})-(\ref{2.2}).

In the cylindrical coordinate system the continuity equation (\ref{2.1}) 
becomes,
\begin{equation}
\label{2.6}
\frac{1}{r}\frac{\partial(ru)}{\partial r} +\frac{\partial w}{\partial z}=0.
\end{equation}
This can be rewritten as
\begin{equation}
\label{2.7}
\frac{1}{r}\left[\frac{\partial(ru)}{\partial r} +\frac{\partial (rw)}{\partial z}\right]=0.
\end{equation}
It follows then that it is appropriate to introduce Stokes stream function 
$\psi$ [22,23] which satisfy
\begin{equation}
\label{2.8}
u=\frac{1}{r}\frac{\partial \psi}{\partial z},\,\,\, w=-\frac{1}{r}\frac{\partial \psi}{\partial r}
\end{equation}
and with these definitions (\ref{2.1}) is satisfied automatically by $\psi$. 
Since $\rho=\rho(r,z)$, (\ref{2.2}) in this frame is
\begin{equation}
\label{2.9}
u\rho_r+w\rho_z=0
\end{equation}
Expressing $u,w$ in terms of $\psi$ we obtain
\begin{equation}
\label{2.10}
J\{\rho,\psi \}=0
\end{equation}
where for any two (smooth) functions $F,G$
\begin{equation}
\label{2.11}
J\{F,G\}=\frac{\partial F}{\partial r}\frac{\partial G}{\partial z} -
        \frac{\partial F}{\partial z}\frac{\partial G}{\partial r}.
\end{equation}
The explicit form of (\ref{2.03}) is
\begin{equation}
\label{2.05}
u\left(\frac{\partial v}{\partial r}+\frac{v}{r}\right)+
w\frac{\partial v}{\partial z}=0.
\end{equation}
From (\ref{2.8}) we infer that (\ref{2.04}) will be satisfied if 
\begin{equation}
\label{2.04}
v=\frac{{\tilde f}(\psi)}{r}
\end{equation}
where ${\tilde f}$ is an arbitrary smooth function of $\psi$. 
However since (\ref{2.10}) implies that $\rho=\rho(\psi)$ and $\psi=\psi(\rho)$ 
we can rewrite (\ref{2.04}) as 
\begin{equation}
\label{2.04a}
v=\frac{f(\rho)}{r}
\end{equation}
We observe that another possible solution of (\ref{2.05}) corresponds to the 
special case where $u(r,z)=0$. Under this restriction $v=v(r)$ remains as an 
arbitrary function of $r$. We shall not consider this possibility in 
this paper.

The momentum equations (\ref{2.02}) in this coordinate system become
\begin{equation}
\label{2.12}
\rho (uu_r +wu_z) = -p_r -\rho\Phi_r +\rho \frac{f(\rho)^2}{r^3}
\end{equation}
\begin{equation}
\label{2.13}
\rho (uw_r+ww_z) = -p_z -\rho\Phi_z,
\end{equation}

To eliminate $p$ from (\ref{2.12}), (\ref{2.13}) we differentiate 
these equations with respect to $z,r$ respectively and subtract. We obtain;
\begin{eqnarray}
\label{2.14}
&&\rho_r(uw_r+ww_z)+\rho(uw_r+ww_z)_r- \\ \notag
&&\rho_z(uu_r+wu_z)-\rho(uu_r+wu_z)_z= \\ \notag
&&-J\{\rho,\Phi\}-J\{\rho,\frac{H(\rho)}{r^2}\},
\end{eqnarray}
where
$$
H(\rho)=\frac{f^2}{2}+\rho f\,f_{\rho}.
$$
For the first and third terms on the left hand side of this equation 
we obtain using (\ref{2.9})
\begin{eqnarray}
\label{2.15}
&&\rho_r(uw_r+ww_z)-\rho_z(uu_r+wu_z) = \\ \notag
&&\rho_r(ww_z+uu_z)-\rho_z(uu_r+ww_r)= \\ \notag
&&J\{\rho,\frac{u^2+w^2}{2}\}
\end{eqnarray}
Similarly for the second and forth terms on the left hand side of (\ref{2.14})
we have
\begin{equation}
\label{2.16}
\rho\left[(uw_r+ww_z)_r-(uu_r+wu_z)_z\right] = \rho\left[u\chi_r + w\chi_z+(u_r+w_z)\chi \right]
\end{equation}
where $\chi=w_r-u_z$. However from (\ref{2.6}) we have
$$
u_r+w_z= -\frac{u}{r}.
$$
Using this equality and expressing $u,w$ in terms of $\psi$ leads to
\begin{equation}
\label{2.17}
u\chi_r + w\chi_z+(u_r+w_z)\chi =-J\{\psi,\frac{\chi}{r}\}.
\end{equation}

Hence we finally obtain that
\begin{equation}
\label{2.18}
\rho\left[(uw_r+ww_z)-(uu_r+wu_z)\right] = \rho J\left\{\psi,\frac{1}{r^2}\left(\nabla^2\psi-\frac{2}{r}\frac{\partial \psi}{\partial r}\right)\right\} .
\end{equation}
Combining the results of (\ref{2.14}),(\ref{2.15}) and (\ref{2.18}) it 
follows that
\begin{eqnarray}
\label{2.19}
&&J\{\rho,\frac{u^2+w^2}{2}\}+ \rho J\left\{\psi,\frac{1}{r^2}\left(\nabla^2\psi-\frac{2}{r}\frac{\partial \psi}{\partial r}\right)\right\}= \\ \notag
&&-J\{\rho,\Phi\}-J\{\rho,\frac{H(\rho)}{r^2}\}.
\end{eqnarray}
To express (\ref{2.19}) in terms of $\rho$ only we use the fact that 
$\psi=\psi(\rho)$ and therefore
\begin{equation}
\label{2.20}
\psi_r =\psi_{\rho}\rho_r,\,\,\ \psi_z =\psi_{\rho}\rho_z,\,\,\
\nabla^2\psi=\psi_{\rho\rho}[\rho_r^2+\rho_z^2] + \psi_{\rho}\nabla^2\rho.
\end{equation}
Using these relations we have 
\begin{equation}
\label{2.21}
J\left\{\rho,\frac{u^2+w^2}{2}\right\} = J\left\{\rho,\frac{\psi_{\rho}^2}{2r^2}(\rho_r^2+\rho_z^2)\right\}
\end{equation}
\begin{equation}
\label{2.22}
\rho J\left\{\psi,\frac{1}{r^2}\left(\nabla^2\psi-\frac{2}{r}\frac{\partial\psi}{\partial r}\right)\right\} = \rho J\left\{\rho,\frac{1}{r^2}\left[\psi_{\rho}^2(\nabla^2\rho -\frac{2}{r}\rho_r) +\psi_{\rho}\psi_{\rho\rho}(\rho_r^2+\rho_z^2)\right]\right\}
\end{equation}
Substituting these results in (\ref{2.19}) leads to
\begin{equation}
\label{2.23}
J\left\{\rho,\frac{\rho}{r^2}\left[\psi_{\rho}^2(\nabla^2\rho -\frac{2}{r}\rho_r) +\psi_{\rho}\psi_{\rho\rho}(\rho_r^2+\rho_z^2)\right] +\frac{\psi_{\rho}^2}{2r^2}(\rho_r^2+\rho_z^2) +\Phi +\frac{H(\rho)}{r^2}\right\} = 0
\end{equation}
This implies that
\begin{equation}
\label{2.24}
\frac{\rho}{r^2}\left[\psi_{\rho}^2(\nabla^2\rho -\frac{2}{r}\rho_r) +\psi_{\rho}\psi_{\rho\rho}(\rho_r^2+\rho_z^2)\right] +\frac{\psi_{\rho}^2}{2r^2}(\rho_r^2+\rho_z^2) +\Phi+\frac{H(\rho)}{r^2} = S(\rho)
\end{equation}
where $S(\rho)$ is some function of $\rho$.

Introducing,
$$
h(\rho)=\rho\psi_{\rho}^2,\,\,\, h^{\prime}(\rho)=\frac{dh(\rho)}{d\rho}.
$$
We can rewrite (\ref{2.24}) more succinctly
\begin{equation}
\label{2.25}
h(\rho)\left(\nabla^2\rho -\frac{2}{r}\rho_r \right) +
\frac{h^{\prime}(\rho)}{2}(\rho_r^2+\rho_z^2) = r^2\left( S(\rho) -
\Phi-\frac{H(\rho)}{r^2}\right)
\end{equation}
This can be rewritten in the form
\begin{equation}
\label{2.29}
h(\rho)^{1/2}\nabla {\bf{\cdot}} ( h(\rho)^{1/2}\nabla \rho)-
\frac{2h(\rho)}{r}\rho_r = r^2\left(S(\rho) -\Phi-\frac{H(\rho)}{r^2}\right) .
\end{equation}
Using (\ref{2.4}) we can eliminate $\Phi$ from (\ref{2.29}) \
to obtain one fourth order equation for $\rho$ only;
\begin{equation}
\label{2.30}
\nabla^2\left\{\frac{1}{r^2}\left[h(\rho)^{1/2}\nabla {\bf{\cdot}} ( h(\rho)^{1/2}\nabla \rho)- \frac{2h(\rho)}{r}\rho_r\right] \right\}+
4\pi G\rho= \nabla^2 S(\rho)-\nabla^2\left(\frac{H(\rho}{r^2}\right).
\end{equation}

Thus we reduced the original nonlinear system of
partial differential equations (\ref{2.1})-(\ref{2.4}) to a coupled system
of two second order equations consisting of (\ref{2.4}),(\ref{2.25}) or 
one fourth order equation for $\rho$.

\subsection{The Interpretation of the Functions $S(\rho),\,h(\rho)$}
 
The function $h(\rho)$ can be considered as a 
parameter function which is determined by the momentum (and angular 
momentum) distribution in the fluid. From a practical point of view the 
choice of this function determines the structure of the steady state 
density distribution. The corresponding flow field can be computed then 
aposteriori (that is after solving for $\rho$) from the following relations; 
\begin{equation}
\label{2.26}
u=-\frac{1}{r}\sqrt{\frac{h(\rho)}{\rho}} \frac{\partial \rho}{\partial z}, \,\,\,
w=\frac{1}{r}\sqrt{\frac{h(\rho)}{\rho}} \frac{\partial \rho}{\partial r}.
\end{equation}

The function $S(\rho)$ that appears in (\ref{2.25}) can be determined 
from the asymptotic values of $\rho$ and $\phi$ on the boundaries of the
domain on which eqs. (\ref{2.5}),(\ref{2.25}) are solved. 
When these asymptotic values are imposed or known one can
evaluate the left hand side of (\ref{2.25}) on the domain boundaries
and re-express it in terms of $\rho$ only to determine $S(\rho)$ 
(on the boundary of the domain). However the resulting functional 
relationship of $S$ on $\rho$ must then hold also within the domain 
itself since $S$ does not depend on $r,z$ directly. For example
on an infinite domain let $h(\rho)=1$, and 
\begin{equation}
\label{2.27}
\displaystyle\lim_{r,z \rightarrow \infty} \rho(r,z) = e^{-r^2} ,\,\,\,
\displaystyle\lim_{r,z \rightarrow \infty} \Phi(r,z) =-e^{-r^2}.
\end{equation}
From (\ref{2.25}) we then have asymptotically that  
\begin{equation}
\label{2.28}
S(\rho)=3e^{-r^2}+\frac{H(\rho)}{r^2}= 3\rho-\frac{H(\rho)}{\ln(\rho)}
\end{equation}
When such asymptotic relations are not given, $S(\rho)$ can be viewed as 
a "gauge". In the following we let $S(\rho)=0$ under these circumstances. 

\subsection{The Steady State Pressure}

In order to derive (\ref{2.25}) we eliminated the pressure from equations
(\ref{2.12})-(\ref{2.13}). However in practical astrophysical 
applications it is important to know the equation of state of the fluid
under consideration. For this reason we derive here an equation analogous to
(\ref{2.25}) for the steady state pressure. To this end we divide 
(\ref{2.12})-(\ref{2.13}) by $\rho$, differentiate the first with respect
to $z$ the second with respect to $r$ and subtract. Using (\ref{2.6}) this 
leads to
\begin{equation}
\label{3.1}
-\frac{u}{r}\chi +u\frac{\partial\chi}{\partial r}+ 
w\frac{\partial\chi}{\partial z} = \frac{1}{\rho^2}J\{\rho,p\}
-J\{\rho,\frac{f\,f_{\rho}}{r^2}\}.
\end{equation}
Expressing $u,w$ and $\chi$ in terms of $\psi$ this yields
\begin{equation}
\label{3.2}
\rho^2\,J\left\{\psi,\frac{1}{r^2}\left[\nabla^2 \psi-\frac{2}{r}\frac{\partial\chi}{\partial r}\right] \right\} = J\{\rho,p \}-
\rho^2J\{\rho,\frac{f\,f_{\rho}}{r^2}\}.
\end{equation}
Eliminating $\psi$ from this equation (using (\ref{2.20})) leads to; 
\begin{equation}
\label{3.3}
J\left\{\rho,\frac{1}{r^2}\left[\rho\psi_{\rho}^2(\nabla^2\rho -\frac{2}{r}\rho_r)+\rho\psi_{\rho}\psi_{\rho\rho}(\rho_r^2+\rho_z^2)\right] \right\} 
=\frac{1}{\rho}J\{\rho,p \}-\rho J\{\rho,\frac{f\,f_{\rho}}{r^2}\}.
\end{equation}
Hence 
\begin{equation}
\label{3.4}
h(\rho)\left( \nabla^2 \rho -\frac{2}{r}\rho_r\right)+ \frac{1}{2}\left[h^{\prime}(\rho) - \psi_{\rho}^2 
\right] (\rho_r^2 + \rho_z^2)= r^2\left(\frac{p}{\rho} - 
\frac{\rho f\,f_{\rho}}{r^2}+P(\rho)\right).
\end{equation}
where $P(\rho)$ is some function of $\rho$. 
Subtracting this equation from (\ref{2.25}) we then have
\begin{equation}
\label{3.5}
\frac{p}{\rho}= S(\rho) - P(\rho) - \frac{1}{2r^2} \psi_{\rho}^2 
(\rho_r^2 + \rho_z^2) -\Phi -\frac{f^2}{2r^2}.
\end{equation}
Therefore the solution of (\ref{2.25}) and (\ref{2.5}) determines
the pressure distribution in the fluid (assuming that the functions
$P,S$ have been determined from the boundary conditions). 

Conversely if the pressure distribution is known apriori e.g if we assume that
the fluid is a polytropic gas where $p=A\rho^{\alpha+1}$ then (\ref{2.4})
can be used to eliminate $\Phi$ from (\ref{3.5}). 
\begin{equation}
\label{3.6}
\nabla^2(P) = \nabla^2\left[S-A\rho^{\alpha}- \frac{1}{2r^2} \psi_{\rho}^2
(\rho_r^2 + \rho_z^2) \right] - 4\pi G \rho-
\nabla^2\left(\frac{f^2}{2r^2}\right)
\end{equation}
It follows then that for a polytropic gas eqs. (\ref{3.4}),(\ref{3.6})
form a closed system of coupled equations for $\rho$ and $P$ with 
a parameter function $\psi_{\rho}^2$. However if we eliminate $P$ from
these two equations we recover (\ref{2.30}). 

As an example for the determination of $P(\rho)$ consider a polytropic
star where the asymptotic behavior of $\rho$ and $\Phi$ are given by
(\ref{2.27}). Substituting these expressions in (\ref{3.5}) we find that
\begin{equation}
\label{3.7}
P(\rho)=4\rho-A\rho^{\alpha}-2\psi_{\rho}^2\rho^2
-\frac{\rho f f_{\rho}}{r^2}
\end{equation}

\setcounter{equation}{0}

\section{The special case where $f(\rho)$ is constant}

When $f(\rho)$ is a constant (or can be approximated 
by a constant) we set $H(\rho)=\Omega^2$ and (\ref{2.30}) becomes
\begin{equation}
\label{2.30a}
\nabla^2\left\{\frac{1}{r^2}\left[h(\rho)^{1/2}\nabla {\bf{\cdot}} ( h(\rho)^{1/2}\nabla \rho)- \frac{2h(\rho)}{r}\rho_r\right] \right\}+
4\pi G\rho= \nabla^2 S(\rho)-\frac{4\Omega^2}{r^4}.
\end{equation}

\subsection{Solutions for (\ref{2.30a}) with $h=1$}

Eq. (\ref{2.30a}) is, in general, a nonlinear equation which (to our 
best knowledge) can not be solved (in general) analytically. The only 
exception is the case where $h$ is a constant under which the resulting 
equation is linear. It should be remembered however that although (\ref{2.30a}) 
reduces to a linear equation when $h$ is a constant the original equations 
(\ref{2.1})-(\ref{2.4}) of the model are nonlinear for this choice of
$h$ as is evident from (\ref{2.26}). Therefore, in principle,
are still attempting to solve to a system of nonlinear equations.

For this choice of $h$,\,we have from (\ref{2.26}) that  
$$
(u,w) = \frac{1}{r\sqrt{\rho}}\left(\frac{\partial \rho}{\partial z},
\frac{\partial \rho}{\partial r}\right).
$$ 
That is with the same gradient of $\rho$, $(u,w)$ 
will increase as $\rho$ decreases. We conclude then that, in general, 
matter in regions with low density might have higher momentum than in 
regions of higher density.  
(In the following we let $S(\rho)=0$.)

With $h=1$ in (\ref{2.30}) this equation takes the following form
\begin{equation}
\label{4.1}
r^3[\frac{\partial^4 \rho}{\partial r^4}+\frac{\partial^4 \rho}{\partial z^4}+
2\frac{\partial^4 \rho}{\partial z^2\partial r^2}]
-4r^2[\frac{\partial^3 \rho}{\partial r^3}+\frac{\partial^3 \rho}{\partial z^2\partial r}]+r[9\frac{\partial^2 \rho}{\partial r^2} +4\frac{\partial^2 \rho}{\partial z^2}] -9\frac{\partial \rho}{\partial r}+4\pi G r^5\rho =-\Omega^2r
\end{equation}
It is possible to find analytic solutions for (\ref{4.1}) using 
first order perturbation expansion in $G$. We consider two strategies.

\subsection{Solutions for (\ref{4.1}) with $\rho=\rho(r)$}

Under this restriction (\ref{4.1}) reduces to,
\begin{equation}
\label{4.49}
r^3\frac{d^4\rho}{dr^4}-4r^2\frac{d^3\rho}{dr^3}+
9r\frac{d^2\rho}{dr^2}-9\frac{d\rho}{dr}+ 4\pi G r^5\rho=-\Omega^2r.
\end{equation}
To solve this equation we assume that $ 4\pi G r^5\rho \ll 1$ and
use first order perturbation expansion in $G$ viz.
$\rho=\rho_0(r)+G\rho_1(r)$.
The resulting equation for $\rho_0$ is
\begin{equation}
\label{4.50}
r^3\frac{d^4\rho_0}{dr^4}-4r^2\frac{d^3\rho_0}{dr^3}+
9r\frac{d^2\rho_0}{dr^2}-9\frac{d\rho_0}{dr}=-\Omega^2r.
\end{equation}
A particular solution of this equation is
\begin{equation}
\label{4.51p}
\rho_p=-\frac{1}{16}\Omega^2r^2(1+2\ln(r)).
\end{equation}
The general solution for the homogeneous part of (\ref{4.50}) is
\begin{equation}
\label{4.51}
\rho_h=C_1+C_2 r^2+r^4(C_3+ C_4\ln(r)),
\end{equation}
where $C_i$ $i=1\ldots 4$ are constants.(Hence $\rho_0=\rho_p+\rho_h$)
 
The equation for $\rho_1$ is
\begin{equation}
\label{4.52}
r^3\frac{d^4\rho_1}{dr^4}-4r^2\frac{d^3\rho_1}{dr^3}+
9r\frac{d^2\rho_1}{dr^2}-9\frac{d\rho_1}{dr}=-4\pi r^5\rho_0.
\end{equation}
Substituting the expression for $\rho_0=\rho_p+\rho_h$ 
in (\ref{4.52}) we find that 
\begin{eqnarray}
\label{4.53}
&&\rho_1=-\frac{\pi r^{10}}{86400}\left(120C_4\ln(r)-67C_4+120C_3\right)-
\frac{C_1\pi r^6}{24} \\ \notag
&&\frac{\pi r^8}{36864}(24\Omega^2\ln(r)-7\Omega^2-192C_2)+
C_7 r^4(4\ln(r)-1)+C_6 r^4+C_5 r^2+C_8
\end{eqnarray}
where $C_i$ are constants.

\subsection{Solution by Separation of Variables}.
To find an approximate particular solution $Y_p$ for (\ref{4.1}) we assume 
that $G\ll 1$ and attempt to find $Y_p$ in the form
$Y_p=\rho_p+Gs(r)$. Substituting this expression of $Y_p$ in (\ref{4.1})
and solving for $s(r)$ we obtain to first order in $G$ that
$$
s(r)=\frac{\pi\Omega^2 r^{8}}{36864}(24\ln(r)-7)
$$

To find solutions for the homogeneous part of (\ref{4.1}) (up to superposition) 
we let $\rho=f(r)g(z)$. (This function $f(r)$ should not be confused
with the function $f(\rho)$ used in previous sections). This leads to
\begin{equation}
\label{4.2}
r^3\frac{d^4 g}{dz^4}+[2r^3\frac{d^2 f}{dr^2}-
4r^2\frac{df}{dr}+4rf]\frac{d^2 g}{dz^2}+
[r^3\frac{d^4 f}{dr^4}-4r^2\frac{d^3 f}{dr^3}+
9r\frac{d^2 f}{dr^2} -9\frac{df}{dr}+4\pi G r^5f]g=0
\end{equation}

A separation of variables for this equation is possible in the following 
three cases:
$$
1.\, g(z)=Cz+D.\,\,\, 2. g(z)=Ae^{\lambda z} +Be^{-\lambda z},\,\,\, 
3.\, g(z)=Ecos(kz+\phi).
$$
In all three cases an approximate analytic solution can be obtained using
first order perturbations with $G$ as a small parameter i.e. we set
\begin{equation}
\label{4.5}
f(r)=f_0(r)+ G f_1(r).
\end{equation}
\begin{enumerate}
\item When $g(z)=Cz+D$ the equation for $f(r)$  is 
\begin{equation}
\label{4.5a}
r^3\frac{d^4 f}{dr^4}-4r^2\frac{d^3 f}{dr^3}+
9r\frac{d^2 f}{dr^2} -9\frac{df}{dr}+4\pi G r^5f=0
\end{equation}
This is the same as the homogeneous part of (\ref{4.49}). Hence $f_0=\rho_h$
The equation for $f_1$ is therefore
\begin{equation}
\label{4.52a}
r^3\frac{d^4f_1}{dr^4}-4r^2\frac{d^3f_1}{dr^3}+
9r\frac{d^2f_1}{dr^2}-9\frac{df_1}{dr}=-4\pi r^5\rho_h.
\end{equation}
The solution of this equation is
\begin{eqnarray}
\label{4.53a}
&&f_1=\frac{1}{720}\left[C_4\left(\frac{67}{120}-\ln(r)\right)-C_3\right]\pi r^{10}
-\frac{C_2\pi}{192}r^8-\frac{C_1\pi}{24}r^6+\\ \notag
&&\frac{1}{16}\left[4C_6+C_7(4\ln(r)-1)\right]r^4+\frac{C_5}{2}r^2+C_8
\end{eqnarray}

Hence the general solution for $\rho$ (to first order in $G$) in this case is 
\begin{equation}
\label{4.3a}
\rho=(\rho_h+Gf_1)(Cz+D)+Y_p
\end{equation}

\item When $g(z)=Ae^{\lambda z} +Be^{-\lambda z}$ the equation for $f(r)$ is
\begin{equation}
\label{4.3}
r^3\frac{d^4 f}{dr^4}-4r^2\frac{d^3 f}{dr^3}+
(9r+2\lambda^2 r^3)\frac{d^2 f}{dr^2}
-(4\lambda^2 r^2+9)\frac{df}{dr} + r(4\pi Gr^4 +\lambda^4 r^2+4\lambda^2)f=0.
\end{equation}
We find that the general solution for $f_0$ in this case is
\begin{equation}
\label{4.8}
f_0(r)=(c_1 r^3+c_2 r)Y_1(\lambda r)+(c_3 r^3+c_4 r)J_1(\lambda r)
\end{equation}
where $J_1,Y_1$ are Bessel function of order one of the first and
second kind. The general solution for $f_1(r)$ is obtained by standard
variation of coefficients method and contains integrals of Bessel functions.
This general solution demonstrates that under proper combination of the 
solution coefficients the mass density might oscillates within the star. 

\item The equation for $f$ in the third case is
\begin{equation}
\label{4.4}
r^3\frac{d^4 f}{dr^4}-4r^2\frac{d^3 f}{dr^3}+(9r-2k^2 r^3)\frac{d^2 f}{dr^2}
+(4k^2 r^2 -9)\frac{df}{dr} + r(4\pi Gr^4 +k^4 r^2-4k^2)f=0.
\end{equation}
This equation has solutions in terms of Bessel functions as in the previous 
case.
\end{enumerate}

\subsection{Steady State Solutions for $h=1+G\rho$, $G\ll 1$}

To derive solutions for $h(\rho)=1+G\rho$, with 
$G \rho \ll 1$ we make the approximation  
$$
\sqrt{1+G \rho} \approx 1+\frac{G}{2}\rho
$$ 
in (\ref{2.30}). Furthermore we assume that $\rho(r,z)=\rho(r)$. 
and let $\rho(r)=\rho_0(r)+G\rho_1(r)$. The resulting equation for $\rho_0(r)$
is the same as (\ref{4.50}) and it follows that the general solution
for $\rho_0$ is $\rho_h+\rho_p$. To first
order in $G$ the resulting equation for $\rho_1$ is
\begin{eqnarray}
\label{6.6}
&&4r^2\frac{d^3\rho_1}{dr^3}-9r\frac{d^2\rho_1}{dr^2}+9\frac{d\rho_1}{dr}=
\left(7\frac{d\rho_0}{dr}+3r^3\frac{d^3\rho_0}{dr^3}-
9r^2\frac{d^2\rho_0}{dr^2}-9\rho_0\right)\frac{d\rho_0}{dr}+ \\ \notag
&&\left(r^3\frac{d^4\rho_0}{dr^4}-4r^2\frac{d^3\rho_0}{dr^3}+
9\frac{d\rho_0}{dr}+4\pi r^5\right)\rho_0+
2r^3\left(\frac{d^2\rho_0}{dr^2}\right)^2
\end{eqnarray}
For $\rho_0=\rho_p$ the general solution for $\rho_1$ is
\begin{eqnarray}
\label{6.7}
\rho_1&=&-\frac{\Omega^4r^4}{2048}(8\ln(r)^2-12\ln(r)+7)+
\frac{\pi\Omega^2r^8}{36864}(24\ln(r)-7)+ \\ \notag
&&(C_2-C_3+4C_3\ln(r))r^4+C_1r^2+C_4
\end{eqnarray}
We observe that this solution contains terms with $\Omega^4$ while the 
solution with $h=1$ contained only terms with $\Omega^2$. A lengthy 
(but analytic) solution can be obtained also when we substitute 
$\rho_0=\rho_h+\rho_p$ in (\ref{6.6}) and solve for $\rho_1$.

\subsection{On the Shape of a Rotating Star}

In this section we consider the shape of a rotating star using the 
solution derived in Sec. $4.3$ for $\rho$ for the case where 
$g(z)=Cz+D$. We note that that this solution for
$\rho$ was derived under the assumption that $h=1$ and therefore 
$\psi_{\rho}^2= 1/\rho$. Also (following convention) we impose on the 
pressure the boundary condition $p=0$ (See Ref. [21] p. $54$ and p. $121$).
To compute the pressure we use (\ref{3.4}) with $S(\rho)=P(\rho)=0$.
Under these assumptions the explicit expression for $p$ is 

\begin{equation}
\label{6.37}
p=\frac{\rho}{r^2}\nabla^2\rho-
\frac{1}{2r^2}\left[\left(\frac{\partial\rho}{\partial r}\right)^2+
\left(\frac{\partial\rho}{\partial z}\right)^2\right]-
\frac{2\rho}{r^3}\frac{\partial\rho}{\partial r}
\end{equation}

When $g(z)=Cz+D$ the approximate expression for $\rho$ (to first order in 
$G \ll 1$) is given by (\ref{4.3a}). We substitute this expression in
(\ref{6.37}) and neglect terms with powers of $G$ greater than one. 
To simplify (algebraically) the expression for $p$ further we let 
$C_1=C_2=C_3=C_4=C_6=C_7=0$ and set $D=5$, $G=0.01$, $\Omega=0.125$,
$C_6=-0.01.$, and $C_8=-0.1$. For a star where $z=1$ at $r=0$
and $r=1.2$ at $z=0$ we obtain Fig. 1 for $z=z(r)$. 

This example is representative for the shape of a rotating star.

\setcounter{equation}{0}
\section{Reduction in Spherical Coordinates}

In spherical coordinates we introduce the standard inertial frame
\begin{eqnarray}
\label{7.1}
&&{\bf e}_r = (\sin\phi\cos\theta,\sin\phi\sin\theta,\cos\phi),\,\,
{\bf e}_{\phi} = (\cos\phi\cos\theta,\cos\phi\sin\theta,-\sin\phi)\\ \notag
&&{\bf e}_{\theta} =(-\sin\theta,\cos\theta,0).
\end{eqnarray}
(Observe that in spherical coordinates $r$ stands for the length of
the radius vector. This should not lead to a confusion as the treatment
of the problem in cylindrical and spherical coordinates is separate). 

In this frame we have under present assumptions the 
following expression for the flow,
\begin{equation}
\label{7.3}
{\bf v} = u(r,\phi){\bf e}_r + v(r,\phi){\bf e}_{\phi}+w(r,\phi){\bf e}_{\theta}
\end{equation}
The continuity equation (\ref{2.1}) is
\begin{equation}
\label{7.5}
\frac{\partial u}{\partial r}+\frac{1}{r}\frac{\partial v}{\partial \phi}+
\frac{2u+\cot\phi\, v}{r}=0.
\end{equation}
This can be rewritten as
\begin{equation}
\label{7.6}
\frac{1}{r^2\sin\phi}\left[\frac{\partial(r^2\sin\phi\,u)}{\partial r}+
\frac{\partial (r\sin\phi\,v)}{\partial \phi}\right] =0.
\end{equation}
Hence if we introduce Stokes stream function $\psi$ which is defined
by the relations
\begin{equation}
\label{7.7}
u = -\frac{1}{r^2\sin\phi}\frac{\partial \psi}{\partial \phi},\,\,\,
v= \frac{1}{r\sin\phi}\frac{\partial \psi}{\partial r}.
\end{equation}
then (\ref{2.1}) is satisfied automatically. Similarly (\ref{2.2}) takes 
the following form
\begin{equation}
\label{7.8}
u\frac{\partial \rho}{\partial r}+ 
\frac{v}{r}\frac{\partial \rho}{\partial \phi} =0.
\end{equation}
This can be rewritten in terms of $\psi$ as
\begin{equation}
\label{7.9}
\frac{1}{r^2\sin\phi}J\{\psi,\rho\}=0.
\end{equation}
where (in this context)
\begin{equation}
\label{7.10}
J\{f,g\}=\frac{\partial f}{\partial r}\frac{\partial g}{\partial \phi}-
\frac{\partial f}{\partial \phi}\frac{\partial g}{\partial r}
\end{equation}

From (\ref{7.9}) we infer that $\rho=\rho(\psi)$ or $\psi=\psi(\rho)$.

The explicit expression for the momentum equations (\ref{2.3}) in this 
coordinate system is
\begin{equation}
\label{7.11}
\rho\left(u u_r+\frac{v}{r}u_{\phi}-\frac{v^2+w^2}{r}\right)
=-\frac{\partial p}{\partial r}-\rho\frac{\partial\Phi}{\partial r}
\end{equation}
\begin{equation}
\label{7.12}
\rho\left(u v_r+\frac{v}{r}v_{\phi}-\frac{w^2\cot\phi-uv}{r}\right)
=-\frac{1}{r}\frac{\partial p}{\partial \phi}-
\frac{\rho}{r}\frac{\partial\Phi}{\partial \phi}.
\end{equation}
\begin{equation}
\label{7.13}
u\left(\frac{w}{r}+\frac{\partial w}{\partial r}\right)+
v\left(\frac{w\,\cot\phi)}{r}+
\frac{1}{r}\frac{\partial w}{\partial \phi}\right) =0
\end{equation}
Using (\ref{7.7}) we find that the general solution of (\ref{7.13}) is 
\begin{equation}
\label{7.13a}
w=\frac{\bar{f}(\psi)}{r\sin\phi}
\end{equation}
where ${\bar f}(\psi)$ is any (smooth) function of $\psi$. However
since $\psi=\psi(\rho)$ we can rewrite (\ref{7.13a}) as
\begin{equation}
\label{7.13b}
w=\frac{f(\rho)}{r\sin\phi}
\end{equation}

To eliminate the pressure term from (\ref{7.11}) and (\ref{7.12}) 
we multiply (\ref{7.12}) by $r$ and differentiate with respect to $r$,
then differentiate (\ref{7.11}) with respect to $\phi$ and subtract.
We obtain
\begin{eqnarray}
\label{7.14}
&&r\rho_r\left(uv_r+\frac{v}{r}v_{\phi}\right)+
r\rho\left(uv_r+\frac{v}{r}v_{\phi}\right)_r+\rho\left(uv_r+\frac{v}{r}v_{\phi}\right) \\ \notag
&&-\rho_r(w^2\cot\phi-uv)-\rho(w^2\cot\phi-uv)_r \\ \notag
&&-\rho_{\phi}\left(uu_r+\frac{v}{r}u_{\phi}\right)-\rho\left(uu_r+
\frac{v}{r}u_{\phi}\right)_{\phi}+ 
\rho_{\phi}\left(\frac{v^2+w^2}{r}\right)+\rho\left(\frac{v^2+w^2}{r}\right)_{\phi}= -J\{\rho,\Phi\}.
\end{eqnarray}
Using (\ref{7.8}) we have for the first and sixth terms on the left hand side of
(\ref{7.14}), 
\begin{equation}
\label{7.15}
r\rho_r\left(uv_r+\frac{v}{r}v_{\phi}\right)=
J\{\rho,\frac{v^2}{2}\}
\end{equation}
\begin{equation}
\label{7.16}
-\rho_{\phi}\left(uu_r+\frac{v}{r}u_{\phi}\right)=
J\{\rho,\frac{u^2}{2}\}.
\end{equation}
The second, third, and seventh terms in (\ref{7.14}) can be expressed as
$$
\rho\left\{u\xi_r+\frac{v}{r}\xi_{\phi}+(u_r+\frac{v_{\phi}}{r})\xi\right\}
$$
where $\xi=rv_r-u_{\phi}$. Using (\ref{7.6}) to express $u_r+\frac{v_{\phi}}{r}$ and 
(\ref{7.7}) to express ${u,\,v}$ in terms of $\psi$ it follows that the sum
of these terms can be rewritten as 
$$
\rho J\left\{\psi,\frac{\xi}{r^2\sin\phi}\right\}=
\rho \psi_{\rho}J\left\{\rho,\frac{\xi}{r^2\sin\phi}\right\}
$$ 
Using (\ref{7.7}),(\ref{7.13b}) and (\ref{7.9}) we can express the sum of
the fourth and eighth terms in (\ref{7.14}) as
$$
J\left\{\rho,\frac{f(\rho)^2}{2r^2\sin^2\phi}\right\}.
$$
Finally the sum of the fifth and ninth terms in (\ref{7.14}) can be expressed as
$$
\rho J\left\{\psi,\frac{\psi_r}{r^3\sin^2\phi}\right\}+
\rho f_{\rho}f J\left\{\rho,\frac{1}{r^2\sin^2\phi}\right\}
$$
which can be rewritten in terms of $\rho$ as
$$
\rho \psi_{\rho}^2J\left\{\rho,\frac{\rho_r}{r^3\sin^2\phi}\right\}+
\rho f_{\rho}f J\left\{\rho,\frac{1}{r^2\sin^2\phi}\right\}
$$

Combining all these results (\ref{7.14}) becomes
\begin{eqnarray}
\label{7.17}
&&J\left\{\rho,\frac{u^2+v^2}{2}\right\}+
\rho\psi_{\rho} J\left\{\rho,\frac{\xi}{r^2\sin\phi}\right\}+
J\left\{\rho,\frac{f(\rho)^2}{2r^2\sin^2\phi}\right\}+ \\ \notag
&&\rho \psi_{\rho}^2J\left\{\rho,\frac{\rho_r}{r^3\sin^2\phi}\right\}+
\rho f_{\rho}f J\left\{\rho,\frac{1}{r^2\sin^2\phi}\right\}=-J\{\rho,\Phi\}
\end{eqnarray}

To express this equation in terms of $\rho$ we note that
\begin{equation}
\label{7.21}
\psi_r =\psi_{\rho}\rho_r,\,\,\ \psi_{\phi} =\psi_{\rho}\rho_{\phi},\,\,\
\end{equation}
Using these relations and (\ref{7.7}) we have
\begin{equation}
\label{7.22}
J\{\rho,\frac{u^2+v^2}{2}\}=J\{\rho,\frac{\psi_{\rho}^2}{2r^2\sin^2\phi}
(\rho_r^2+\frac{1}{r^2}\rho_{\phi}^2)\}.
\end{equation}
Expressing $u,\,v$ in terms of $\psi$ we obtain the following representation
for $\xi$
\begin{equation}
\label{7.23}
\sin\phi\xi=\nabla^2 \psi -\frac{3}{r}\frac{\partial\psi}{\partial r}-
\frac{2\cot\phi}{r^2}\frac{\partial\psi}{\partial \phi}.
\end{equation}
But using (\ref{7.21}) we have,
\begin{equation}
\label{7.24}
\nabla^2 \psi =\psi_{\rho}\nabla^2\rho+\psi_{\rho\rho}\left(\rho_r^2+
\frac{1}{r^2}\rho_{\phi}^2\right).
\end{equation}
Hence (\ref{7.17}) can be written as
\begin{eqnarray}
\label{7.25}
&&J\left\{\rho,\frac{\psi_{\rho}^2}{2r^2\sin^2\phi}
(\rho_r^2+\frac{1}{r^2}\rho_{\phi}^2)\right\}+ \\ \notag
&&\rho\psi_{\rho}J\left\{\rho,\frac{1}{r^2sin^2\phi}\left[\psi_{\rho}\nabla^2\rho+
\psi_{\rho\rho}\left(\rho_r^2+\frac{1}{r^2}\rho_{\phi}^2\right)-
\frac{3}{r}\psi_{\rho}\rho_r-\frac{2\cot\phi}{r^2}\psi_{\rho}\rho_{\phi}\right]\right\}+ \\ \notag
&&J\left\{\rho,\frac{f(\rho)^2}{2r^2\sin^2\phi}\right\}+
\rho \psi_{\rho}^2J\left\{\rho,\frac{\rho_r}{r^3\sin^2\phi}\right\}+
\rho f_{\rho}f J\left\{\rho,\frac{1}{r^2\sin^2\phi}\right\}=-J\{\rho,\Phi\}
\end{eqnarray}
Combining all the terms in this equation we infer that
\begin{eqnarray}
\label{7.26}
&&\rho\psi_{\rho}^2\left[\nabla^2\rho-
\frac{2}{r}\rho_r-\frac{2\cot\phi}{r^2}\rho_{\phi}\right]+ \\ \notag
&&\frac{1}{2}\left(2\rho\psi_{\rho}\psi_{\rho\rho}+\psi_{\rho}^2\right)
\left(\rho_r^2+\frac{1}{r^2}\rho_{\phi}^2\right)+ H(\rho) \\ \notag
&&=r^2\sin^2\phi[-\Phi+R(\rho)]
\end{eqnarray}
where $H(\rho)=\frac{f(\rho)^2}{2}+ \rho ff_{\rho}$ and $R(\rho)$ is some 
function of $\rho$. Introducing
$$
h(\rho)=\rho\psi_{\rho}^2,\,\,\, h^{\prime}=\frac{dh}{d\rho}
$$
we can rewrite (\ref{7.26}) as
\begin{eqnarray}
\label{7.27}
&&h(\rho)\left[\nabla^2\rho-
\frac{2}{r}\rho_r-\frac{2\cot\phi}{r^2}\rho_{\phi}\right]+ 
\frac{h^{\prime}}{2}\left(\rho_r^2+\frac{1}{r^2}\rho_{\phi}^2\right)+H(\rho) \\ \notag
&&=r^2\sin^2\phi\left[-\Phi+R(\rho)\right]
\end{eqnarray}

We observe also that
$$
\psi_{\rho}=\left(\frac{h(\rho)}{\rho}\right)^{1/2}
$$
and therefore from (\ref{7.7})
$$
u=-\frac{1}{r^2\sin\phi}\left(\frac{h(\rho)}{\rho}\right)^{1/2}
\frac{\partial \rho}{\partial \phi},\,\,\,
v=\frac{1}{r\sin\phi}\left(\frac{h(\rho)}{\rho}\right)^{1/2}
\frac{\partial \rho}{\partial r}.
$$

The function $R(\rho)$ can be determined if the asymptotic behavior of
$\rho$ and $\Phi$ is known. For example if $h(\rho)=1$, $H(\rho)=0$ and 
\begin{equation}
\label{7.27a}
\displaystyle\lim_{r \rightarrow \infty} \rho(r,\phi) =e^{-r^2\sin^2\phi} ,\,\,\,
\displaystyle\lim_{r,z \rightarrow \infty} \Phi(r,\phi) =-\frac{2}{r^2}
e^{-r^2\sin^2\phi},
\end{equation}
then it follows from (\ref{7.27} that asymptotically 
\begin{equation}
\label{7.27b}
R(\rho)=4\rho
\end{equation}
When such asymptotic relations are not given, $R(\rho)$ can be viewed as
a "gauge". In the following we let $R(\rho)=0$ under these circumstances.

\subsection{Equation for the Pressure in Spherical Coordinates}

To obtain an equation for the pressure in spherical coordinates we divide
(\ref{7.11}) by $\rho$ and differentiate with respect to $\phi$. Similarly 
we multiply (\ref{7.12}) by $r$, divide by $\rho$ and differentiate with
respect to $r$. Subtracting the first result from the second we obtain
\begin{equation}
\label{7.28}
\left\{u\xi_r+\frac{v}{r}\xi_{\phi}+(u_r+\frac{v_{\phi}}{r})\xi\right\}
+\psi_{\rho}^2J\left\{\rho,\frac{\rho_r}{r^3\sin^2\phi}\right\}+
f_{\rho}f J\left\{\rho,\frac{1}{r^2\sin^2\phi}\right\}=
\frac{1}{\rho^2}J\{\rho,p\}
\end{equation}
Using (\ref{7.6}) to express $u_r+\frac{v_{\phi}}{r}$ and
(\ref{7.7}) to express ${u,\,v}$ in terms of $\psi$ it follows that   
we can rewrite (\ref{7.28})
\begin{equation}
\label{7.29}
J\left\{\psi,\frac{\xi}{r^2\sin\phi}\right\}
+\psi_{\rho}^2J\left\{\rho,\frac{\rho_r}{r^3\sin^2\phi}\right\}+
f_{\rho}f J\left\{\rho,\frac{1}{r^2\sin^2\phi}\right\}=
\frac{1}{\rho^2}J\{\rho,p\}
\end{equation}
Following the same procedure used previously we obtain
\begin{eqnarray}
\label{7.30}
&&\rho\psi_{\rho}J\left\{\rho,\frac{1}{r^2sin^2\phi}
\left[\psi_{\rho}\nabla^2\rho+
\psi_{\rho\rho}\left(\rho_r^2+\frac{1}{r^2}\rho_{\phi}^2\right)
-\frac{2}{r}\psi_{\rho}\rho_r-\frac{2\cot\phi}{r^2}\psi_{\rho}\rho_{\phi}
\right]\right\} + \\ \notag
&&\rho f_{\rho}f J\left\{\rho,\frac{1}{r^2\sin^2\phi}\right\}=
\frac{1}{\rho}J\{\rho,p\}
\end{eqnarray}
Hence it follows that
\begin{eqnarray}
\label{7.31}
&&\rho\psi_{\rho}^2\left[\nabla^2\rho
-\frac{2}{r}\rho_r-\frac{2\cot\phi}{r^2}\rho_{\phi}
\right]+\rho\psi_{\rho}\psi_{\rho\rho} \left(\rho_r^2+
\frac{1}{r^2}\rho_{\phi}^2\right)+\rho ff_{\rho}= \\ \notag
&&r^2\sin^2\phi\left(\frac{p}{\rho}+ P(\rho)\right)
\end{eqnarray}
where $P(\rho)$ is some function of $\rho$.

Using $h(\rho)$ we can express (\ref{7.31}) as
\begin{eqnarray}
\label{7.32}
&&h(\rho)\left[\nabla^2\rho
-\frac{2}{r}\rho_r-\frac{2\cot\phi}{r^2}\rho_{\phi}
\right]+\frac{h^{\prime}-\psi_{\rho}^2}{2}\left(\rho_r^2+
\frac{1}{r^2}\rho_{\phi}^2\right)+\rho ff_{\rho}= \\ \notag
&&r^2\sin^2\phi\left(\frac{p}{\rho}+ P(\rho)\right).
\end{eqnarray}
Subtracting (\ref{7.32}) from (\ref{7.27}) yields,
\begin{equation}
\label{7.33}
\frac{p}{\rho}=R(\rho)-P(\rho)+\frac{f^2}{2}-\Phi-
\frac{\psi_{\rho}^2(\rho_r^2+\rho_{\phi}^2)}{2r^2\sin^2\phi}.
\end{equation}
For a polytropic star where $p=A\rho^{\alpha+1}$ we therefore have,
\begin{equation}
\label{7.34}
P(\rho)=R(\rho)-A\rho^{\alpha}+\frac{f^2}{2}-\Phi-
\frac{\psi_{\rho}^2(\rho_r^2+\rho_{\phi}^2)}{2r^2\sin^2\phi}.
\end{equation} 

\section{Summary and Conclusions}

In this paper we considered the steady state Euler-Poisson equations 
with rotations under the additional assumption of density stratification. 
The governing equations of this model consist of six nonlinear partial 
differential equations.
We showed however that this set of equations can be reduced (in cylindrical 
and spherical coordinates) to two. We derived also a separate equation for the 
pressure in the star with special consideration for those stars composed of 
a polytropic fluid. 

Several (analytic) approximate steady state solutions of the model equations 
were obtained. Using these solutions and the boundary condition $p=0$ 
it is possible to derive expressions for the shape of a rotating star. 
An explicit generic example was presented. 

The present paper did not explore numerical solutions of the model equations.
However this can be done with relative ease due to the reduction in the 
number of the model equations. The solution of the reduced model equations 
(\ref{2.1})-(\ref{2.4}) in spherical coordinates has to be carried 
out numerically.

This paper does not provide a general solution to the original
star model described by Euler-Poisson equations. However it does provide
insights and solutions for a subclass of stars described by this model.

\newpage
\section*{References}

\begin{itemize}

\item[1] Auchmuty, J.F.G.; Beals, R.: Variational solutions of some 
nonlinear free boundary problems, Arch. Ration. Mech. Anal. 43, 
pp.255-271, 1971.

\item[2] Beskin V. S.: Axisymmetric stationary flows in compact 
astrophysical objects, Physics-Uspekhi 40 pp. 659-688, 1997.

\item[3] Chandrasekhar, S.: An Introduction to the Study of Stellar Structures. 
University of Chicago Press, Chicago, 1938.

\item[4] Chandrasekhar, S: Ellipsoidal figures of equilibrium-an historical 
account, Comm. Pure Appl. Math, 20 251-265, 1967.

\item[5] Friedman, A., Turkington, B.: The oblateness of an axisymmetric 
rotating fluid, Indiana Univ. Math fluid J., 29 pp.777-792, 1980.

\item[6] Humi M.: Steady States of self gravitating incompressible
fluid. J. Math. Phys. 47, 093101 (10 pages), 2006.

\item[7] Humi M.: A Model for Pattern Formation Under Gravity, 
Applied Mathematical Modelling 40, pp. 41-49, 2016

\item[8] Humi M :Patterns Formation in a Self-Gravitating Isentropic Gas
             Earth Moon Planets https://doi.org/10.1007/s11038-017-9512-y

\item[9] Imamura J. N., Durisen R.H., Pickett B.K.: Nonaxisymmetric 
Dynamic Instabilities of Rotating Polytropes, Astrophysical Journal, 
528,pp. 946-964, 2000.

\item[10] J. Jang J. and Makino T.:
On Slowly Rotating Axisymmetric Solutions of Euler-Poisson Equations
Arch. Ration. Mech. Anal., 225 pp. 873-900, 2017.

\item[11] M. Kiguchi M., S. Narita S., Miyama S. M., Hayashi C.: 
The Equilibria of Rotating Isothermal Clouds, Astrophysical J., 
317 pp.830-845, 1987.

\item[12] Kovetz, A.: Slowly rotating polytropes. Astrophys. J. 154, 
pp. 999-1003, 1968.

\item[13] Kunzle H.P., Nester J.M.: Hamiltonian formulation of 
gravitating perfect fluids and the Newtonian limit, J. Math. Phys 25, 
pp. 1009-1018, 1984.

\item[14] Letelier, P.S., Oliveira, S.R.:  Exact self-gravitating 
disks and rings: A solitonic approach ,J. Math. Phys. 28 pp.165-170, 1987.

\item[15] Li, Y.Y.: On uniformly rotating stars, Arch. Ration. Mech. Anal,
115 pp.367-393, 1991.

\item[16] Luo T and Smoller J.:
Existence and Non-linear Stability of Rotating Star Solutions of the 
Compressible Euler-Poisson Equations, Arch. Ration. Mech. Anal. 
191, pp.447-496, 2009.

\item[17] Matsumoto, T. and Hanawa T.: Bar and Disk Formation 
in Gravitationally Collapsing Clouds. Astrophys. J., 521(2), pp.659-670,1999.

\item[18] Milne, E.A : The equilibrium of a rotating star. Mon. Not. R. 
Astron. Soc. 83, pp.118-147, 1923.

\item[19] Ortega V. G., Volkov E. and Monte-Lima I.:  Axisymmetric 
instabilities in gravitating disks with mass spectrum, Astronomy and 
Astrophysics 366, pp.276-280, 2001.

\item[20] Prentice, A. J. R., (1978) Origin of the solar system,
Earth Moon and Planets, 19, pp. 341-398.

\item[21] Roxburgh I. W., Non-Uniformly Rotating, Self-Gravitating,
Compressible Masses with Internal Meridian Circulation,
Astrophysics and Space Science, 27,pp.425-435, 1974

\item[22] Tassoul Jean-Louis: Theory of Rotating Stars, 
Princeton U press, Princeton, NJ. 1978,

\item[23] Yih C-S: Stratified flows. Academic Press, New York,
NY, 1980.

\end{itemize}

\newpage
\begin{figure}[ht!]
\includegraphics[scale=1,height=160mm,angle=0,width=180mm]{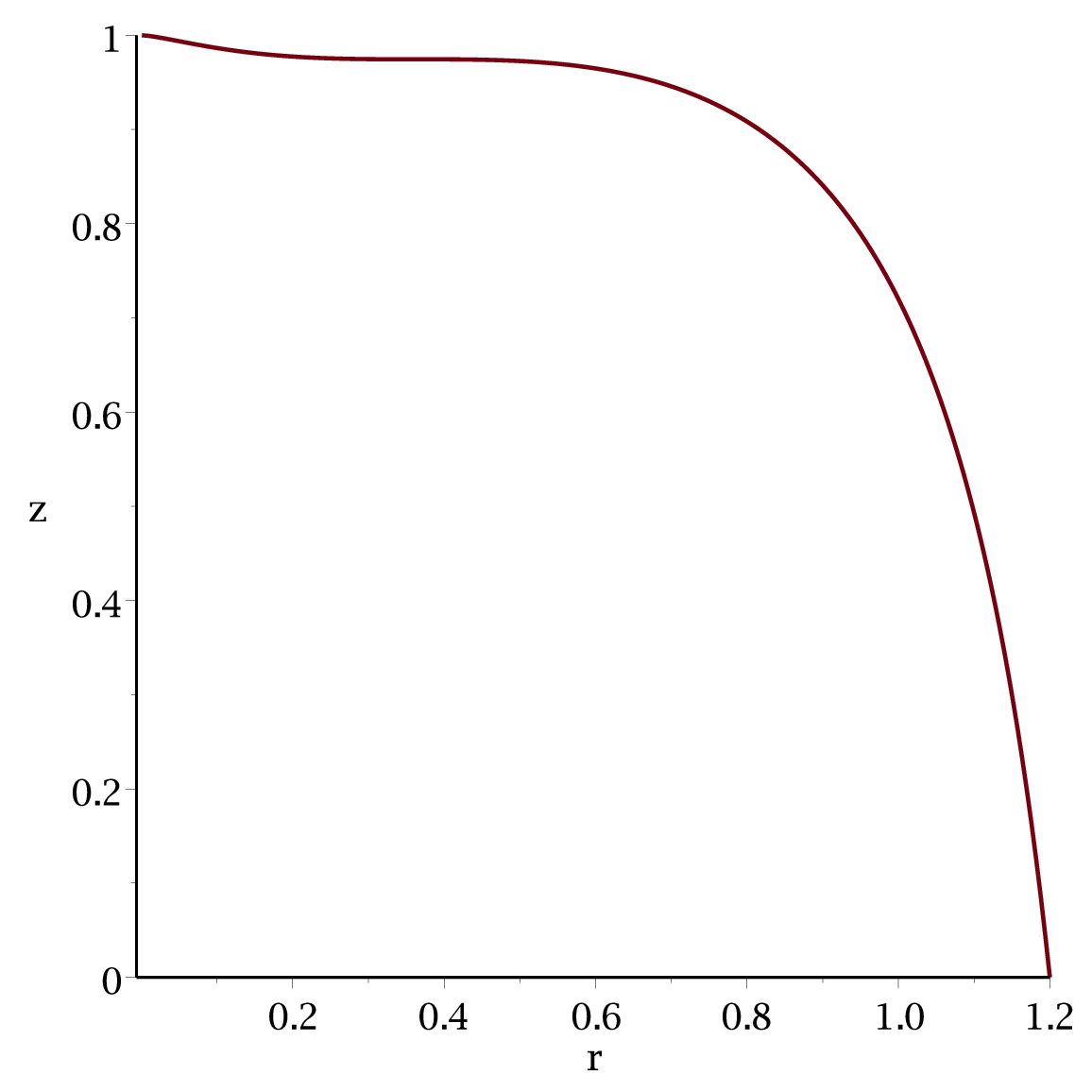}
\label{Figure 1}
\caption{z as a function of r}
\end{figure}

\end{document}